# Introduction of the
# Residue Number Arithmetic Logic Unit
# With Brief Computational Complexity Analysis
## (Rez-9 soft processor)

White Paper
By
Eric B. Olsen

Excerpted from
"General Arithmetic in Residues"
By Eric B. Olsen
(Release TBD)

Original release in confidence
11/12/2012

Rev 1.1 (12/5/12)
Rev 1.2 (1/22/14)
Rev 1.3 (5/27/14)
Rev 1.4 (11/29/15)
Rev 1.45(12/2/15)

Condensed Abstract:

Digital System Research has pioneered the mathematics and design for a new class of computing machine using residue numbers. Unlike prior art, the new breakthrough provides methods and apparatus for general purpose computation using several new residue based fractional representations. The result is that fractional arithmetic may be performed without carry. Additionally, fractional operations such as addition, subtraction and multiplication of a fraction by an integer occur in a single clock period, regardless of word size. Fractional multiplication is of the order $O(p)$, where $p$ equals the number of residues. More significantly, complex operations, such as sum of products, may be performed in an extended format, where fractional products are performed and summed using single clock instructions, regardless of word width, and where a normalization operation with an execution time of the order $O(p)$ is performed as a final step.

## Table of Contents

# Foreword (revised)

Eric B. Olsen

Residue Number Arithmetic Logic Unit (RNS ALU)

Digital System Research Inc has pioneered the mathematics and hardware design for a new class of computing machine using residue numbers. Unlike prior art, the new breakthrough provides methods and apparatus for general purpose computation using several new residue based fractional representations. The result is that fractional arithmetic may be performed without carry. Additionally, fractional operations such as addition, subtraction and multiplication of a fraction by an integer occur in a single clock period, regardless of word size. Fractional multiplication is of the order $O(p)$, where $p$ equals the number of residue digits. More significantly, complex operations, such as sum of products, may be performed in an extended format, where fractional products are performed and summed using single clock instructions, regardless of word width, and where a normalization operation with an execution time of the order $O(p)$ is performed as a final step.

A computing machine implementing the new fractional residue representation will surpass the performance levels of binary at some specific word width and for certain applications. This performance increase will occur both as a result of eliminating carry propagation, and as a result of increased efficiency of calculation due to the underlying mathematics and properties of the new residue number system. The applications and benefits of the new computational method appear numerous. For one, the new residue ALU may be implemented using standard digital hardware, as a binary coded residue computer. Additionally, optical and quantum computing may benefit by adopting this new form of computation, provided such a system support 64 to 128 distinct states. In terms of power consumption, the power efficiency of the new form of calculation may prove advantageous, since each digit operates in relative isolation to each another, and the growth of circuit area due to look ahead carry is eliminated.

Still other advantages may exist; calculations performed using the new number system appears to be quite accurate. The new form of fractional representation supports a minimum of $2^p - 1$ distinct denominators, where $p$ is the number of modulus associated with the fractional range. In comparison, the binary system supports only $n$ distinct denominators for $n$ bits of binary fractional range. The significant increase in the number of distinct denominators of residue fractions promotes the exact representation of more commonly used ratios, and this appears to increase the accuracy of certain residue calculations over binary floating point calculations of comparable word size.

Many applications using the new form of residue computation exist. Increasing word size affects computation speed of the new residue ALU less dramatically then with a binary ALU. Applications for very wide word RNS processors include factorization, scientific computation, cryptography and simulation. Given a non-binary machine capable of higher efficiency due to fundamental mathematical operations, such a machine will displace binary in at least some critical computation tasks.

Since its inception, DSR has aggressively pursued development of its residue based CPU technology. In that time, DSR proved and verified its fundamental technology with its Rez-1 ALU. As a result of the Rez-1 effort, DSR refined its residue processor architecture, discovering many improvements and features which benefit the technology. With this new perspective, DSR started the development of its second generation residue ALU called Rez-9. The Rez-9 co-processor is a scalable residue processor deployed as a soft FPGA based product. Rez-9 supports very wide data formats, yet fits easily into an Altera Cyclone-IV series FPGA. More capable variations of the Rez-9 co-processor target the Altera Stratix-IV series FPGA device, and provide larger data width and faster operation. For more information see www.digitalsystemresearch.com.

## Notice to the Reader

The author has received some complaints regarding this paper. Most complaints involve the paper's lack of explanation regarding how residue fractions are represented and computed. This paper does not treat this subject in any detail. The purpose of this paper is to provide a brief overview of a general purpose residue ALU. The paper also puts forth a mathematical based comparison of the speed of general purpose residue arithmetic versus binary arithmetic. If the reader is interested in apparatus that is capable of performing general purpose residue arithmetic, and the detailed descriptions, definitions and flow charts that underlie its operation, please see reference [6].

## Background

As far back as the late 1950's, computer scientists and researchers have analyzed, proposed and adapted residue number arithmetic to specific computing problems [1], [4]. However, such efforts have been primarily restricted to problems requiring only integer arithmetic. In most cases, the arithmetic has been restricted to addition, subtraction and multiplication. Moreover, prior art computing applications using residue numbers have failed to compete with binary computers in essentially all cases. The reasons are numerous, and include problems related to signed residue arithmetic, residue number comparison, residue number conversion and the perception that residue numbers are only integers [1], [2], [3].

Digital System Research (DSR) has solved the problems of processing residue numbers confronted in the prior art. DSR has pioneered a new form of fractional representation, as well as the underlying operations which enable general purpose, signed, fractional computation in residue number format. The developments from DSR constitute a new method and apparatus for general purpose arithmetic computation; they constitute a fundamentally new and different approach to performing basic arithmetic. Therefore, it is not only the hardware ALU which has been invented, but a new form of arithmetic calculation which is mathematically different than binary.

## What is a residue number ALU?

The term "residue ALU" is relatively new. DSR defines a residue number ALU (RNS ALU) to be an arithmetic logic unit which 1) uses signed integer and fractional residue number representation as its primary number formats, and 2) is capable of performing general purpose arithmetic computation using residue numbers alone.

The residue number system is not a fixed radix system, and does not have the same number of digit states for each digit. The number of digit states may be large; for example, the Rez-9 ALU can have digits with as many as 509 states (Q=9). Therefore, the RNS ALU is not restricted to having only a zero or one digit as in binary. However, in order to build a residue number ALU using digital electronics, we must encode digit states using binary, so strictly speaking we are building a "binary coded residue number ALU".

Using binary to encode another number system is not new. For example, one might recall the use of "binary coded decimal", commonly denoted BCD. In order to be useful, the binary coded number system must offer an advantage. In the prior art, the advantage is generally not speed, since straight binary encoding is generally faster and more memory efficient than any number system encoded in binary. One exception to this is with residue numbers. Residue numbers do not operate using carry, and therefore, it is possible to gain efficiencies using the residue number system that lead to faster execution versus pure binary encoding.

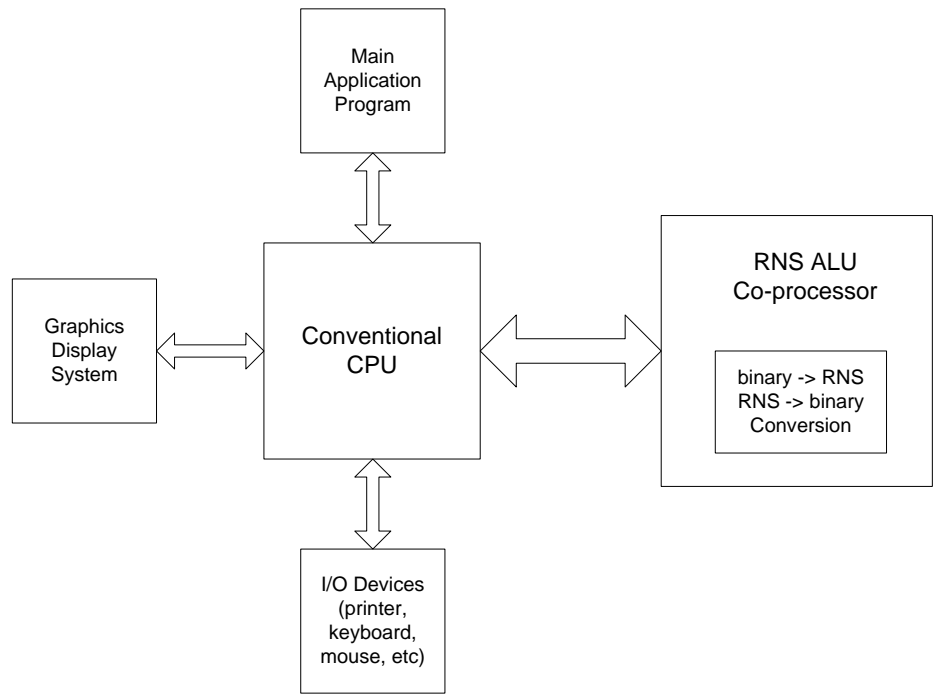

RNS ALU Co-processor
Figure 1

## Why hasn't an RNS ALU been developed before?

Computer designers have contemplated and even constructed residue number ALU prototypes in the past. See references [1], [2], [3], [4]. These prior art systems have demonstrated integer residue math, or at most, a scaled integer math. However, this paper makes a bold statement; there is no known prior art residue computer system designed for general purpose calculation. Using the inventions put forth by DSR, we can now achieve practical, general purpose calculation using residue numbers.

A simple block diagram of a typical RNS computer system is shown in figure 1. One key to the invention is the development of efficient fractional residue number representations [6]. In other words, DSR has invented an equivalent to a fixed point or floating point number but in residue format. As seen in figure 1, within the RNS ALU, DSR has developed the means to process real numbers (similar to floating point types) and perform meaningful and useful calculation entirely and continuously in residue format.

The long sought after dream of capitalizing on the carry free properties of the residue number system have been achieved, both in theory and in practice by DSR.

## What is a residue ALU good for?

DSR is working on establishing a complete answer to this question. However, there are some strong indications that have emerged. For one, the residue ALU will have an advantage over binary for very wide word processing. This is because the speed of basic residue operations such as addition and integer multiplication is not decreased by an increase in ALU bit width. For binary, this is not true.

On the other hand, multiplication of a fractional residue value with another fractional residue value executes in about the same time as a digit based binary multiplier. Both multipliers require more time if the number of digits increases.

However, an amazing result is that the operation of product sums is much more efficient using the RNS ALU. The reason is that products are computed using integer math, and then summed in their extended format, where the final sum is normalized using an operation similar to a single fractional multiply. This means matrix processing of wide words is a perfect fit for the RNS ALU. This is a major discovery from the research labs of DSR.

## How do you use the residue ALU?

As a result of breakthroughs at DSR, it is now clear residue numbers can represent fractional quantities as accurately or more accurately then binary. However, there are also distinct differences. For one, there is no concept of carry using residue numbers. Secondly, the residue number system is not "weighted". This is to say that a residue number format does not allow one to easily distinguish the value of a number, nor does it allow one to easily generate an action, such as by use of a D/A converter.

We refer to binary as a "weighted" system, whereas residue numbers are considered a "non-weighted" number system. Residue numbers must be converted to a weighted number system in order for the data to be used. Before processing starts, initial data may exist in binary and be converted to residue format prior to use by the residue ALU. However, initial data may also be generated within the residue ALU as well.

Once in residue format, the data is processed and tested by the ALU to perform the necessary arithmetic and logic operations. Even though the residue data is in a non-weighted format, it is directly and correctly processed by the residue ALU. Once the result of the processing is complete, the residue data may be converted back to binary using a high speed hardware conversion apparatus, or alternatively, a slow software based solution.

Therefore, we introduce another series of key inventions by DSR; they are high speed conversion apparatus for the new residue ALU. The new conversion apparatus can be diagrammed as four converters, although in practice the converters may share resources. The four converters are 1) forward integer converter, 2) forward fractional converter, 3) reverse integer converter, and 4) reverse fractional converter. The introduction of fractional number conversion between binary and RNS is a new concept introduced by DSR.

Figure 2 illustrates a residue ALU combined with a series of high speed converter apparatus. We refer to this combination as a residue processor unit, or RPU. The RPU is a new invention by DSR, which is illustrated by the use of unique fractional number converters, and the development of a processing unit that processes data in a purely residue domain. All required arithmetic processing is performed entirely in residue number format until the resultant data is ready to be converted back to binary, and used in the real world, i.e., to plot an image on a graphics display, for example.

It is typical that the RPU be under some control, such as by direct instruction control by the host CPU. Additionally, the RPU may execute its own instructions directly. In either case, the RPU acts much like a floating point unit in that it is designed to process arithmetic calculations; therefore, one application of the RPU is as a math co-processor to a standard binary CPU. This arrangement is shown in figure 2.

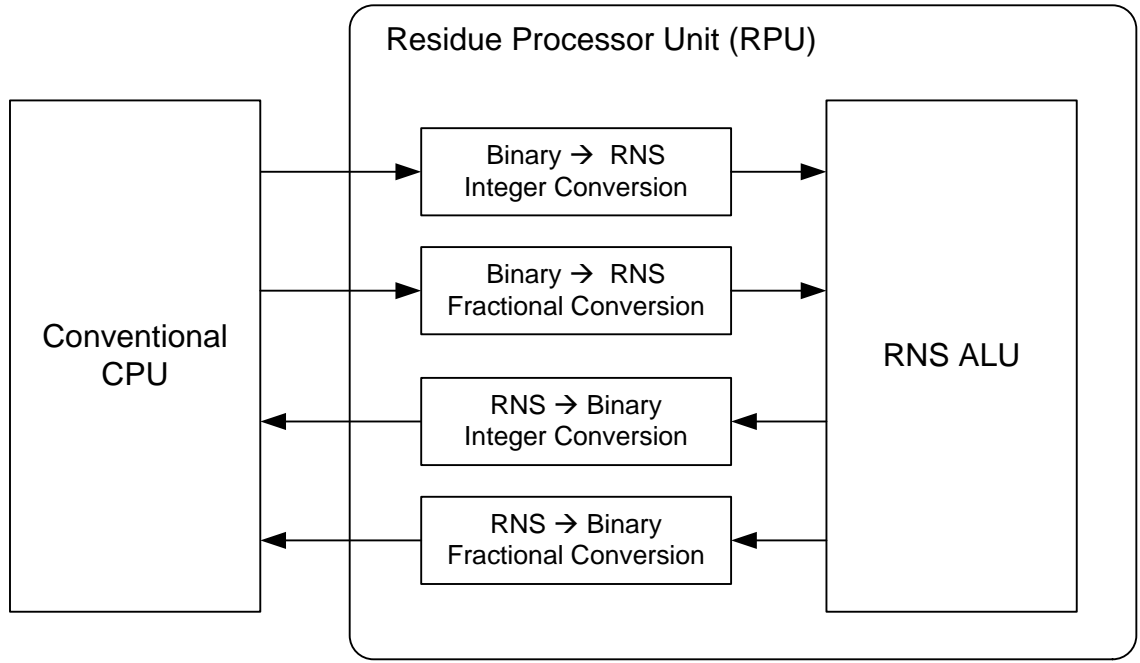

High Speed Integer and Fractional Residue
Conversion Apparatus from DSR
Figure 2.

Other traditional RNS architectures are also possible, such as architectures common in DSP applications. In these cases, high speed hardware apparatus from DSR may be used to perform the conversion between any word size binary and residue word formats required of the application. With the breakthroughs at DSR, it is now possible to process any DSP algorithm using fractional residue mathematics, and so any known DSP function may be easily constructed using this new method of machine computation.

## Computational complexity analysis:

The computational complexity of an algorithm is often described using the big O notation, or Landau notation [5]. This notation is used to describe some specified behavior or measure of the algorithm as some variable tends to infinity. In terms of performance of an arithmetic algorithm, it is common to describe the relative time of execution as a function of growing word size. For example, the time of execution of a circuit or algorithm is measured as the word width of the operands is increased.

Computational time complexity of arithmetic operations is important to establish, and is an on-going goal of DSR. It is also a key methodology for providing results and drawing conclusions. However, comparison of systems using stated computational complexity can also lead to faulty conclusions. There must be care in interpreting such stated complexity analysis claims, since in practice actual performance is influenced by a wide variety of well known reasons. Established complexity analysis ratings have built in perspective, and stated or unstated assumptions, meaning that one must be careful to compare the same things in the same perspective, and under the same conditions.

For example, in the literature one can find the speed of binary multiplication anywhere from $O(\log(n))$ to the $O(n^2)$. In each case, special assumptions apply. For example, the fast asymptotic speed of $O(\log(n))$ is claimed from the perspective of IC designers, who are willing to trade a square increase in circuit area for doubling word size, or who assume ever denser logic functions of ever increasing logic inputs. On the other extreme are arbitrary precision software libraries and software routines, which may operate as slow as $O(n^2)$, since the software approach performs operations on numbers in a digit by digit fashion, and commonly resorts to using inefficient, "schoolbook" calculations.

Some standards for asymptotic run time of various arithmetic operations have been established and tabulated by Wikipedia [5]. Many of these relations reflect software approaches to multiplication, i.e., algorithms using a single processor to manipulate values digit by digit. For example, the execution of multiplication using the Karatsuba algorithm shows an execution time complexity of $O(n^{1.585})$, where $(n)$ represents the width in bits of the operands. Alternatively, the elementary "schoolbook" algorithm has an execution time of $O(n^2)$, which is quite poor. This is not unrealistic since it may be necessary to break up a value into digits in order to process the value in practice, especially if $(n)$ gets very large. This is the type of comparison we will be making here.

But what is the correct order of execution for hardware multiplication using binary? There isn't an easy answer. As mentioned, the execution time of binary multiplication as a function of word size is stated from many viewpoints. For an extendable binary multiplier we suggest a bit linear run time, which yields an $O(n)$ for multiplication, i.e., by using a shift, carry-store, and add hardware solution. For our fractional residue multiplier, we have a digit linear run time; that is, the run time is proportional to the number of residue digits $(p)$. Therefore, run time comparison is performed by plotting the growth of residue digits versus effective bit width in Table 1. While this comparison is not exactly fair, it provides useful insight.

In the second analysis, we upgrade the binary multiplier to use the same size digit width as the residue ALU. If we sub-divide the binary word size $(n)$ into Q-bit wide digits, and we assume a digit linear run time, we can state an $O(n/Q)$ execution time for the binary multiplier. We can then correctly adjust the first analysis, resulting in a new and precise mathematical comparison of the run time comparison between the residue ALU and an equivalent binary ALU, which is shown in Table 4.

- Complexity analysis comparisons are made more difficult when comparing the performance of two ALUs operating with radically different number systems.

Table 1 shows our first published comparison of the RNS ALU execution times versus execution times of common binary algorithms. The relations of Table 1 illustrate the relative increase in execution time as the width of the operands grows in digits. For the binary ALU, the number of digits is the number of bits denoted as (n). For the RNS ALU, the number of digits is denoted as (p).

Table 1 illustrates all of the basic integer and fractional operations, including addition, subtraction, multiplication and division. In the case of addition and multiplication, several alternate execution rates are provided for completeness. Often, the real stated complexity bound is some relation in between the two extremes.

For example, in the first row of Table 1, the stated asymptotic run time of binary addition is simply O(n), where (n) is the number of digits of the word; this relationship assumes a linear increase in execution time as a consequence of handling carry for each digit added. We've included the O(n) execution time for completeness, however, this execution rate is likely too conservative. If we assume that an effective look-ahead carry circuit can be maintained for all increases of operand width, we can assign an estimated time complexity of n/log(n) for binary addition, as shown in the second row of Table 1.

| Complexity Analysis of execution time vs. word width, in (n) bits or (p) digits | | | | |
|---|---|---|---|---|
| **Operation** | **Binary** | | **RNS** | |
| | **Binary Algorithm** | **(n) bits wide** | **RNS Algorithm** | **(p) digits wide** |
| Integer/Fractional Addition | Schoolbook addition | O(n) | basic | Constant |
| | Look-ahead carry | n/log(n) | basic | |
| Integer/Fractional Subtract | Look-ahead carry | n/log(n) | basic | Constant |
| Integer Multiplication | Karatsuba algorithm | $O(n^{1.585})$ | basic | Constant |
| | Hardware shift | O(n) | basic | |
| Integer Division | Hardware shift | O(n) | Olsen RNS Integer Division | TBD |
| Fractional Multiplication | | O(n) | Olsen RNS Multiplier | O(p) |
| Fractional Division | Newton-Raphson | ~O(n*logn) | Goldschmidt-Olsen | ~O(p*logp) |
| Matrix multiply | Strassen Algorithm | $O(M^{2.807} * n)$ | Strassen + Olsen product sum | $O(M^2 * p)$ |
| Comparison | | ~ Constant C | known | O(p) |
| Binary to RNS Conversion | | N/A | known | O(n) |
| RNS to Binary Conversion | | N/A | Olsen | O(p) |

**Table 1.**

In either case, the RNS ALU outperforms the binary ALU for addition, since the time complexity is theoretically constant, regardless of word width. The fact that RNS addition execution time is theoretically constant is well known in the prior art. For subtraction, the results are essentially the same as addition. RNS subtraction enjoys a constant execution time regardless of word size, binary subtraction does not.

For multiplication, the story is more complex. For integer multiplication, the RNS ALU execution time is again constant, denoted by "Constant" in Table 1. Generally, RNS addition, subtraction and multiplication of integers require only a single clock, or single arithmetic cycle, since all digits are processed in parallel without carry. For multiplication of fractional quantities, that is, fixed point values, execution is proportional to (p), where p is the number of RNS digits.

Therefore, in the case of the RNS ALU, we must distinguish integer multiplication from fractional multiplication, since integer multiplication has a constant execution time versus (p), and fractional multiplication has a linear execution time versus (p), i.e. O(p). For binary, it does not matter; in either the integer or fractional case, the execution time of a multiply is linear with respect to the operand width (n), or O(n). This is due to carry for both integer and fractional multiplication when using binary.

For the case of residue integer division, DSR does not have an execution rate for its algorithm. However, it is expected to be less desirable than O(p). Integer division has been marked "TBD" until this research is complete. For binary, we have integer division listed as O(n); however, this *may be misleading*, and may be closer to O(n*log(n)) or even O($n^2$) as the operand width (n) increases without bound. For fractional division, the table assumes a Newton-Raphson algorithm, or Goldschmidt type algorithm, so the execution rate of fractional division is approximated as the rate of fractional multiplication times a log factor, since the Newton-Raphson routine exhibits quadratic convergence. Table 4 lists the order of execution of fractional division the same as fractional multiplication for simplicity.

For matrix multiplication, or the process of computing product sums, a stunning result is achieved using residue arithmetic. With the RNS ALU, product terms may be processed in an extended format using integer multiplication, and summed in an extended format using integer addition, where the final sum is normalized using (p) steps similar to a single RNS fractional multiply. This new form of computation essentially transforms a fractional product sum into a series of integer operations and a single fractional normalization; the RNS ALU can perform each of these integer operations in a single clock cycle, regardless of word width. The single fractional normalization occurs in O(p). This is an incredible advancement to scientific processing!

Using Wikipedia, we've restated the complexity order for matrix multiplication ($MxM$) using Strassen Algorithm method as O($M^{2.807}$ * n) for a binary ALU. For the RNS ALU, the result is far better, since product terms are not a function of (p), so the stated result is O($M^2$ * p). The exponent of M is not a typo, since for each dot product, M-1 multiplications are converted to integer multiplications, having an execution O(M * $C_m$), where $C_m$ is a constant and does not grow with (p). In other words, fractional multiplications (and additions) of order O(p) are converted to operations requiring only constant time, or O(1), so an additional factor of M is divided out of the time complexity relation for $MxM$ number of "sum of product" operations. This is a significant breakthrough in computer science.

Another significant advantage of RNS matrix multiplication is that the calculation may be performed and maintained using the full resolution and range of the results. After the final stage of product summation, a final normalization is performed. The normalization performs a final truncation, and a single conditional rounding operation. This new multiplication process decreases processing time yet significantly increases accuracy as well. Not even an array of binary floating point units (FPU) can compensate for this feature, or emulate this feature.

For certain other operations, the binary ALU will outperform the RNS ALU. For example, we are suggesting a constant time for binary number comparison. This is not completely true, but is generally used as a reasonable assumption. The time complexity for RNS value comparison is O(p). These stated order of execution rates are average, and assume arbitrary random numbers are being compared. However, if numbers being compared are approximately equal, then the order of execution for binary may be linear.

The results of RNS processing cannot be used without conversion to binary or some other weighted number system. DSR has developed advanced, high speed conversion apparatus that allows the results of fractional and integer quantities to be converted to binary for use in the real world. We have stated forward conversion time complexity as O(n/Q), n being the number of bits of the binary source. Reverse conversion time complexity is proportional to (p), the number of digits of the RNS ALU. This is fortunate, since combining RNS fractional multiplication with conversion to binary does not increase the time complexity of RNS multiplication.

## Required Corrections to the Complexity Analysis

The story told in Table 1 looks good, maybe too good. As cautioned, one should pay careful attention to the details of any comparison analysis. However, it is instructive to plot and compare the performance of the RNS ALU and the binary ALU, versus (n), the number of bits wide of the operands. In Table 1, execution rates for the residue ALU are given in terms of the number of digits, or (p). Therefore, a conversion is required between the two range domains of each number system for a fair comparison.

There are a number of corrections that should be applied to Table 1 to get a clearer picture of the comparison of the two methods of arithmetic. For one, there is a relationship between (n) and (p) that may be used to establish a common metric for our execution rate table. Furthermore, we adapt a more advanced binary model that compares more reasonably to the execution time model of the RNS ALU. This new binary model is valid, as it models cascading faster and denser binary arithmetic units.

To compare ALU's of the same binary width, we derive the following. Consider the following RNS number system consisting of (p) prime digit modulus:

$$M_1 = 2, \ M_2 = 3, \ M_3 = 5, \ M_4 = 7, \ldots, \ M_p = P = pth\ prime$$

We herein define this basic sequence of RNS modulus as the "natural" residue number system. The range of the natural residue number system is:

$$Range(p) = R(p) = M_1 * M_2 * M_3 * \ldots * Mp$$

When comparing a binary ALU to an RNS ALU, we use (n) to represent the binary ALU width, ($n_e$) to represent the effective binary width of the RNS ALU, and (p) to designate the number of residue digits of the RNS ALU. Therefore, the effective bit width of the RNS ALU range is:

$$n_e = log_2(R(p)) = log_2(2 * 3 * 5 * \ldots * P)$$

Now, the relationship of equivalent binary bits to RNS ALU digits can be established. Since the equivalent bit width $n_e$ and the binary bit width $n$ will be equal for fair comparison, then:

$$n \approx n_e = log_2(R(p))$$

To find our relationship, we need to know the growth rate of our natural residue number system. Thus, we use equation (1a), an approximate and alternate form of the asymptotic growth of primorials [7]:

$$p \approx log(R(p)) / log(p) \qquad (1a)$$

Therefore, the number of natural RNS digits (p) equivalent to a bit width (n) is approximated as:

$$p \approx n / log_2(p) \qquad (1b)$$

Equation 1b is an interesting result since (p) grows more slowly than (n). The reason for slow growth of (p) is the radix of each successive RNS modulus increases, and does not remain fixed as with binary. In other words, the growth of the range of (p) residue digits is not at a fixed exponential rate, it grows even faster with each additional digit. Therefore, when we view (n) for each equivalent (p) in terms of range, we see (n) grows more quickly. Relation (1a) has significance for number theory since the range, or product, of the first (p) primes is closely related to its number (p), i.e. the asymptotic growth of primorial numbers.

These ideas are illustrated in Table 2. The data of Table 2 is essentially driven by the characteristics of the residue number system. For example, once a digit width in bits is chosen (Q), the maximum number of residue digits is then defined in the second column (p). The equivalent number of decimal digits for the maximum digit residue number system is shown in column 3. Equivalent binary width in bits is shown in column 4. The ratio of the number of equivalent bits to the number of RNS digits is shown in the last column. This column clearly shows the growth rate for effective bits is faster than the growth rate for residue digits.

| RNS Digit Width (Q) | Maximum RNS Digits (p) | Effective decimal digits | Equivalent binary width ($n_e$) | RNS / Binary ratio ($p/n_e$) |
|---|---|---|---|---|
| 8 bit | 54 | 101 | 335 | 0.16 |
| 9 bit | 97 | 211 | 703 | 0.14 |
| 10 bit | 172 | 427 | 1420 | 0.12 |
| 11 bit | 309 | 862 | 2865 | 0.11 |
| 12 bit | 564 | 1749 | 5811 | 0.10 |
| 13 bit | 1028 | 3502 | 11635 | 0.09 |
| 14 bit | 1900 | 7059 | 23452 | 0.08 |

**Table 2.**

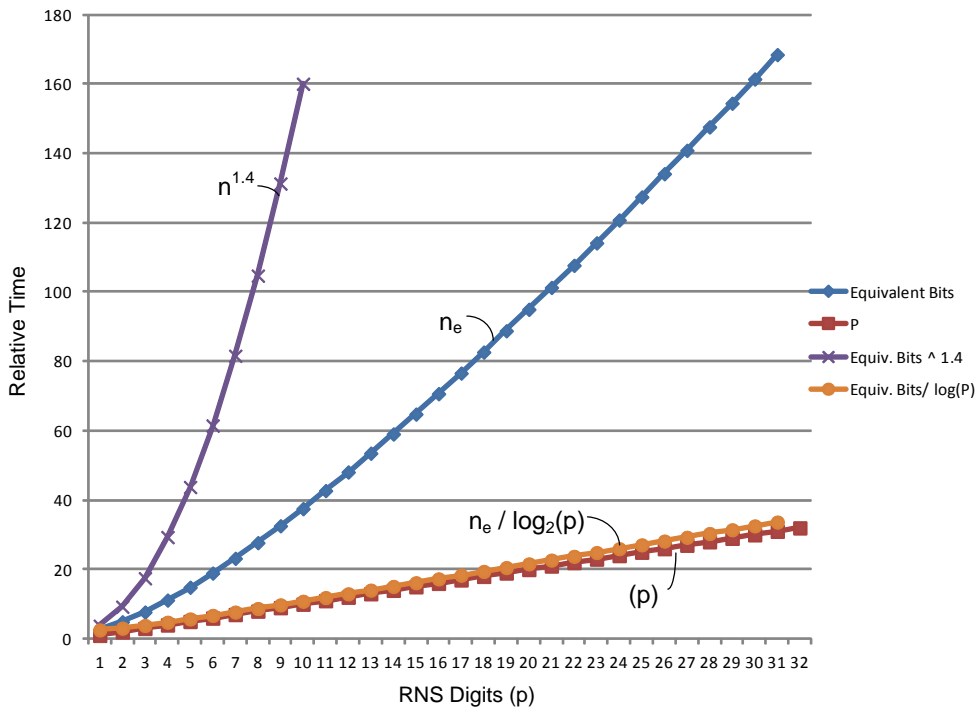

**GRAPH 1.**

The growth of binary bits versus residue digits is also shown using Graph 1. In this case, the comparison is made over a smaller range of residue digits, in this case one to 32 residue digits.

Graph 1 shows (n) versus (p) for the same equivalent binary width. This is the same as comparing the multipliers of the RNS ALU and the "bit linear" binary ALU in our analysis. Graph 1 essentially illustrates execution time versus operand width, with the binary ALU having the ($n_e$) curve, and the residue ALU having the (p) curve. Since the bit width and effective bit width are approximately equal, the RNS multiplier shown by curve (p) is much faster than the bit linear binary multiplier ($n_e$). The term "bit linear" specifies that the binary "digit width" is always a single bit in this chart.

Graph 1 also illustrates that the number of natural residue digits for an equivalent width RNS ALU is (approximately) the number of effective binary bits divided by the logarithm (base two) of the number of RNS digits. This approximation is within 7.2% at an effective width of 335 bits, and remains within 12.6% at an effective width of 23,452 bits; this error, or divergence, is shown as the small space between the bottom two curves of the graph. This divergence is the error of the approximation of equation (1a). Corrections can be applied to relationship (1b) to make it more accurate, but is beyond the scope of this paper.

One can see that if multiplying RNS digits is as fast as multiplying binary bits, the RNS ALU is much faster. Graph 1 also shows a typical rate for the Karatsuba algorithm, which indicates how slow the software algorithm is compared to both hardware approaches.

The curves of graph 1 properly reflect the relationships between (n) and (p), but they do not properly reflect a realistic comparison of binary and RNS multiply algorithms using similar techniques and data width. To do this, we must adopt a more efficient binary digit width, which we will call Q. If we divide the binary word into a plurality of Q bit wide digits, the number of digits of the binary multiplier is approximately (n)/Q. Compared with the bit wise multiplier, the number of digits is now reduced because we are operating on more than one bit at a time at the digit processing level. In fact, mathematically, we may allow Q to change in our relations since Q will change for the RNS ALU as well. For our comparison, the fairest choice for Q is related to our residue number system, since the residue number has a fixed "minimum" digit encoding width, which is dictated by the largest modulus used. Therefore, we choose Q, our digit encoding width in bits, to follow the minimum Q of the natural RNS system:

$$Q = \lfloor Log_2(P) \rfloor + 1 \qquad (2)$$

The value P is the largest digit modulus of the residue number. For example, if the largest modulus is P=61, then Q=6, since six bits is required to store the digits of the modulus=61. Table 3 is provided to show adjusted digit conversion data; this new data is adjusted for equivalent bit width ($n_e$) as well as equivalent digit width, (Q).

In Table 3, a third column showing a revised number of binary "digits" is provided. The number of binary digits is defined as the equivalent bit width divided by Q and rounded up. It can be seen that in this new light, the binary multiplier is no longer a "bit linear" multiplier, but is assumed a "digit linear" multiplier. This means that in terms of equivalent bits, the binary ALU is stated as running faster than O(n). This is justified in comparison, as the residue ALU requires the support of ever increasing digit width arithmetic functions, whose execution time (up to this point) has been assumed to be constant regardless of Q. In other words, the residue digit function, or table look-up, for the digit modulus P=16381 is assumed to be as fast as the digit modulus P=13 in Table 3.

| Digit width (Q) | Max RNS digits (p) | Binary Digits (radix=$2^Q$) | Equivalent binary width ($n_e$) | Largest modulus P |
|---|---|---|---|---|
| 4 | 6 | 4 | 14.87 | 13 |
| 5 | 11 | 8 | 37.55 | 31 |
| 6 | 18 | 13 | 76.63 | 61 |
| 7 | 31 | 24 | 161.46 | 127 |
| 8 | 54 | 42 | 334.88 | 251 |
| 9 | 97 | 79 | 702.60 | 509 |
| 10 | 172 | 142 | 1419.52 | 1021 |
| 11 | 309 | 261 | 2864.48 | 2039 |
| 12 | 564 | 485 | 5810.32 | 4093 |
| 13 | 1028 | 895 | 11634.09 | 8191 |
| 14 | 1900 | 1676 | 23451.13 | 16381 |

**Table 3**

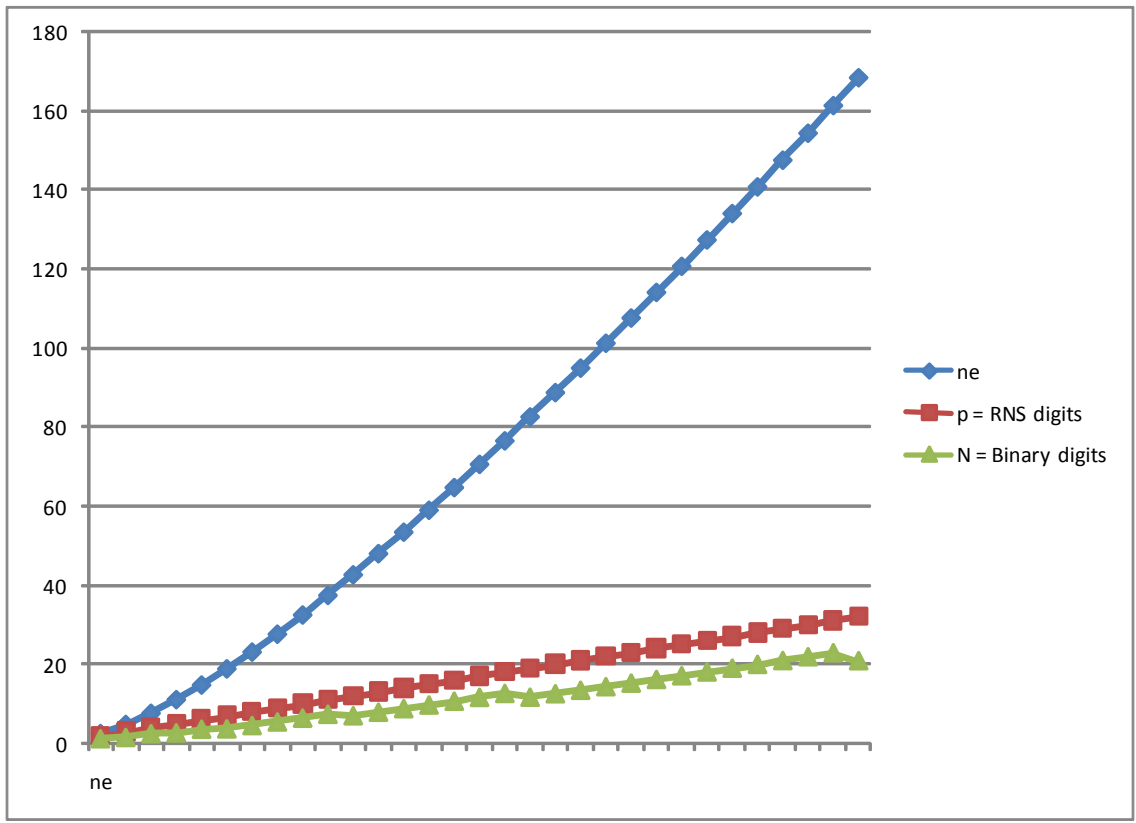

**Graph 2.**

Graph 2 shows the corrected binary multiplier execution time versus the RNS multiplier. As in graph 1, the horizontal axis is the range of the number systems, given in terms of residue digits, using a scale of one to 32 residue digits. The vertical access is effective bit width. The table need only plot the growth rate of digits of each system as the range of the number system is increased. Graph 2 shows that the number of binary digits is less than the number of RNS digits for each equivalent (n). After all, we should not expect the RNS representation to be more efficient than the best case binary representation. In other words, it takes more residue digits to make an equivalent size binary word consisting of the same sized digits. However, it is also noted that the digit growth rates are approximately equal.

Our comparison indicates both multipliers have approximately the same digit growth rate when compared to the bit-wise binary multiplier (n). Table 4 is provided below, and is corrected for equivalent bit width, and equivalent digit width in bits.

Also see graph 2. Using the growth rate in digits of each number system for comparison, the two multipliers are closely matched over a large range. Using digit growth to compare our two multipliers, we see that they have approximately the same growth with respect to the equivalent binary width ($n_e$). While graph 2 shows digit growth rates are roughly equal, we will be more specific in our analysis later.

| Operation | Binary | | RNS | |
|---|---|---|---|---|
| | **Binary Algorithm** | n bits wide | **RNS Algorithm** | $n_e$ bits wide |
| Integer/Fractional Addition | Schoolbook addition | $O(n/Q)$ | basic | Constant |
| | Look-ahead carry | $(n/Q)/\log(n/Q)$ | basic | |
| Integer/Fractional Subtract | Look-ahead carry | $(n/Q)/\log(n/Q)$ | basic | Constant |
| Integer Multiplication | Karatsuba algorithm | $O(n/Q)^{1.585}$ | basic | Constant |
| | Hardware shift | $O(n/Q)$ | basic | |
| Integer Division | Newton-Raphson | $O(n/Q)$ | Olsen Division | TBD |
| Fractional Multiplication | | $O(n/Q)$ | Olsen Multiplier | $O(n/\log_2 p)$ |
| Fractional Division | Newton-Raphson | $O(n/Q)$ | Goldschmidt-Olsen | $O(n/\log_2 p)$ |
| MxM Matrix multiply (delayed normalization) | Standard Algorithm | $O(M^3 * n/Q)$ | Olsen | $O(M^3*(1)+ M^2*(n/\log_2 p))$ |
| | Strassen Algorithm | $O(M^{2.807} * n/Q)$ | Strassen-Olsen | $O(M^{2.807}*(1)+ M^2*(n/\log_2 p))$ |
| Comparison | | ~ Constant C | MRC | $\sim O(n/\log_2 p)$ |
| Binary to RNS Conversion | | N/A | basic | $O(n/Q)$ |
| RNS to Binary Conversion | | N/A | Olsen | $O(n/\log_2 p)$ |

**Time vs. word width (n) - Corrected for same equivalent effective bit width ($n_e$)**

**Table 4. Arithmetic Operations and Functions for n-bit wide operands**

## Interpretation of time complexity analysis

Table 4 results are a more realistic performance comparison of a binary ALU versus the residue based ALU as the effective binary width ($n_e$) is increased. The new comparison accounts for the correct equivalent bit width, which favors the RNS ALU; Table 4 also corrects for the use of Q wide digits in both RNS and binary, which favors the binary multiply. Using the restriction of equation 2, we observe that P, the largest digit modulus, is larger than (p), the number of residue digit modulus; therefore, the number of binary digits (n/Q) is smaller versus (p), for the same effective bit width $n_e$. This establishes the fact that the number of RNS digits is slightly larger than its binary counterpart of the same Q. This might indicate the binary multiplier is better, but this is not the case. There are several reasons in practice, i.e., the binary multiplier will require twice as many clocks as its number of digits, and the binary multiplier may require an additional two's complement operation for negative numbers requiring an additional full carry.

| Q | p | n$_e$ | log(p) | n$_e$/log(p) | log(p)/Q log(p)/log(P) | n$_e$/Q | 2*n$_e$/Q |
|---|---|---|---|---|---|---|---|
| 6 | 18 | 89 | 4.16993 | 21.34 | 0.6950 | 14.83 | 30 |
| 7 | 31 | 183 | 4.95420 | 36.94 | 0.7077 | 26.14 | 52 |
| 8 | 54 | 335 | 5.75489 | 58.21 | 0.7194 | 41.88 | 84 |
| 9 | 97 | 703 | 6.59991 | 106.52 | 0.7333 | 78.11 | 156 |
| 10 | 172 | 1420 | 7.42626 | 191.21 | 0.7426 | 142.00 | 284 |
| 11 | 309 | 2865 | 8.27146 | 346.37 | 0.7520 | 260.45 | 521 |
| 12 | 564 | 5811 | 9.13955 | 635.81 | 0.7616 | 484.25 | 969 |
| 13 | 1028 | 11635 | 10.00562 | 1162.85 | 0.7697 | 895.00 | 1790 |
| 14 | 1900 | 23452 | 10.89178 | 2153.18 | 0.7780 | 1675.14 | 3350 |

Table 5.

To further illustrate and aid in the interpretation of the complexity analysis, Table 5 is provided. In Table 5, actual values governing the two number systems are tabulated for each value of Q, and therefore, each equivalent binary width (n$_e$). The tables mainly show actual digit growth rates of each system in addition to values derived from the theoretical equations developed.

As mentioned earlier, the fractional residue multiplier may be faster than its equivalent binary multiplier in practice. The reason is that integer portions of a fractional number may be separated out and processed rapidly using residue representation. For example, an N (N=n/Q) digit binary multiplier may require 2*N number of clocks in a basic configuration. This data is shown in the last column of Table 5. On the other hand, the fractional residue multiplier can achieve a signed multiply in only (p) number of clocks, shown in the second column of Table 5 for equivalent binary width. For each range shown in Table 5, (p) is less than 2*n/Q. When we combine all of the advantages, the residue ALU may compute both fractional and integer quantities in significantly less time than a binary ALU of equivalent size and equivalent digit architecture.

The conclusions above assume the numeric representation for real numbers has roughly the same precision for both the integer and fractional portions. If we only compare numeric representations having only a fractional part, the residue ALU has no additional benefit, as it requires up to 2*(p) clocks to perform a fractional multiply. Readers interested in improved residue apparatus related to these improvements are referred to reference [8].

Mathematically, we require better proof of the growth rate of (p) residue digits versus (n/Q) binary digits as (n$_e$) grows to infinity. For example, does there come a point where the residue digits more than double the number of binary digits? To answer this, it is noted Q is essentially equal to log$_2$(P), where P is the largest residue modulus. (See equation 2). Therefore, taking log$_2$(P) equal to Q, and using the data from Table 5, one can see that the ratio of log(p) / log(P) is a value that grows slightly but is asymptotic to no greater than a value of one. The efficiency of RNS digits actually increases slightly versus binary digits as (n$_e$) grows to infinity. This ratio is plotted in graph 3.

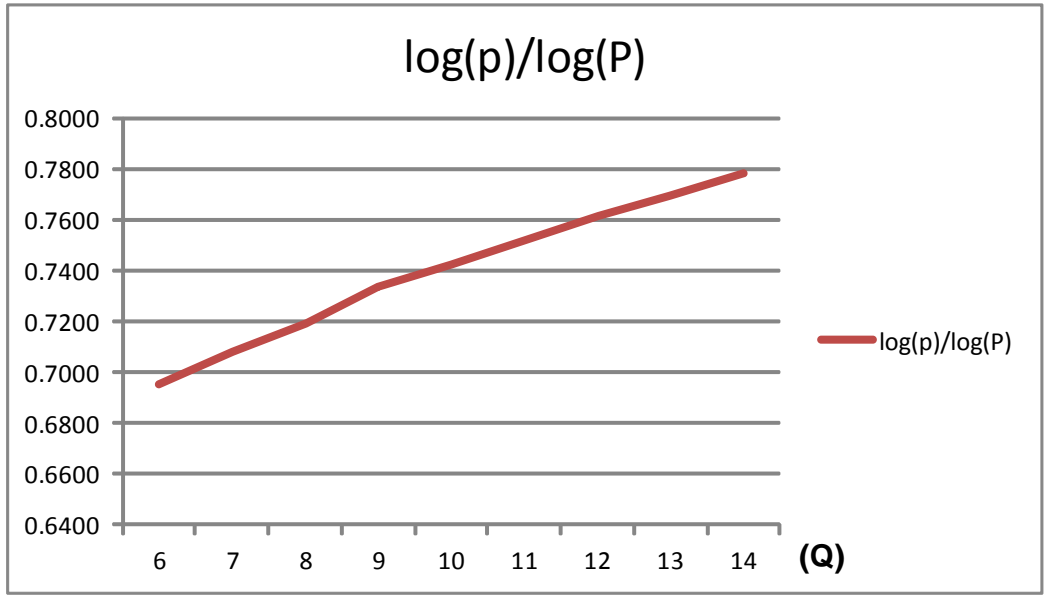

Graph 3.

## Conclusion of time complexity analysis

In conclusion, it may appear somewhat odd that attributes of the natural residue number system, namely (p) and (P), do indeed provide the theoretical basis for the comparison of the digit based RNS and binary multipliers. The growth of the number of primes in a natural sequence, versus the growth of the largest prime in that sequence, aptly describes the growth rate of the digits in the two number systems of fair comparison. This growth characterizes the performance comparison of the residue versus the binary multiplier of equivalent digit shift architecture. If the binary digit factor is multiplied by two, as needed to account for a practical multiplier apparatus, then the residue multiplier may be faster in practice.

The mathematical treatment provided herein is worst case when performing a fair comparison of similar digit based binary and RNS (fractional) multipliers. In practice, an RNS design may choose its modulus more carefully so as to maximize the representational efficiency. In addition, a practical RNS ALU may use powers of the smaller prime modulus to essentially fill-out the Q number of bits implicitly allocated for each respective digit, see reference [6]. Even for applications such as cryptography, large primes which are relatively close to the value $2^Q$ may be appropriately chosen for the design of digit modulus. In other words, the number of residue digits (p) and the number of optimum binary digits (n/Q) is often equal or at least very close. However, our math shows that (n/Q) is always less than or equal to (n/log p) in all cases.

Despite being the conclusion section, we'll introduce the equation for RNS representational efficiency, $E_R$, for interested readers:

$$E_R = \frac{log_2(M_1 * M_2 * ... M_p)}{p*Q} \; x \; 100\% \tag{3}$$

Equation 3 is applicable to any system which uses a series of modulus $M_1$ through $M_p$, and a fixed digit encoding width of Q bits. For example, the Rez-9 ALU has a 9 bit digit representation (Q=9), and supports a representational efficiency of over 95%. This means the difference in number of digits between the Rez-9 and a comparable digit based binary multiplier is about +5%.

Furthermore, the results of RNS ALU matrix multiplication are nothing less than dramatic, and are surprisingly independent of the run time comparison of multiplication alone. The processing of fractional product sums using a residue ALU is more efficient than using a binary ALU. However, not considered until now is the time required for matrix data conversion. The execution rate of reverse conversion of an *MxM* matrix is $O(M^2 * n/\log_2(p))$, therefore the order of conversion and multiplying a matrix is equal with respect to (p), and when combined is faster in most cases than the $O(M^{2.807} * n/Q)$ rate of the binary ALU execution alone. Forward conversion is $O(M^2 * n/Q)$, which is faster than reverse conversion. Thus, there is strong mathematical incentive to pursue RNS processing when the application requires repetitive processing of product sums, such as iterative matrix processing using high precision data. One finds such iterative matrix processing is common in many scientific applications.

The final comparison results of Table 4 are astonishing for other reasons. The RNS ALU can perform an integer multiply in a single clock, or a multiply of an integer by a fractional quantity in a single clock, or an addition or subtraction of a fractional quantity in a single clock, all regardless of effective bit width. Furthermore, the RNS ALU performs with superior representational accuracy. Given a specific application involving enough calculations that are more efficient in RNS than binary, the RNS ALU may have a solid advantage, and therefore find practical use.

It should be noted that the residue based ALU is new, and in its infancy. It is anticipated that new techniques and innovations will improve the efficiency of the RNS operations. Furthermore, new techniques of applying the new form of calculation will lead to better overall optimizations for end applications.

### Other important metrics for the RNS ALU

Speed and performance comparison with conventional systems are important, but are not the only relevant factors. The new ALU demonstrates many other potential benefits. For example, the IC circuit topology may be easier to manage and arrange as the number of digits is increased since there is no increase in the complexity of carry circuits.

As stated previously, the RNS ALU shows promise in applications requiring digit extendable, wide word ALUs and CPUs. Additionally, the new RNS ALU shows advantages in terms of required logic resources, circuit area, and power dissipation. Furthermore, the new ALU shows increased accuracy, as a result of supporting significantly more denominators in its fractional representation. Many of these parameters must be compared with respect to an increasing number of bits of resolution, since it is along this axis that the RNS ALU excels. For example, the residue ALU may show advantages in speed, circuit area, power dissipation and numerical accuracy as the effective ALU accumulator width, $(n_e)$, increases.

Despite the underlying technology and implementation, we anticipate certain measures of the residue ALU will surpass its binary counterpart at some word width and for some applications. Therefore, one important goal at DSR is to determine the applications which the RNS ALU will be advantageous.

# Bibliography, related & references cited